\documentclass{llncs}

\usepackage[T1]{fontenc}
\usepackage[dvips]{graphicx}
\usepackage{subfigure}
\begin{document}
\title{Modeling Stop-and-Go Waves in Pedestrian Dynamics}
\author{Andrea Portz\inst{1} and Armin Seyfried\inst{1}}

\institute{J\"ulich Supercomputing Centre, Forschungszentrum J\"ulich,
  52425 J\"ulich, Germany}

\maketitle

\begin{abstract}
Several spatially continuous pedestrian dynamics models have been
validated against empirical data. We try to reproduce the
experimental fundamental diagram (velocity versus density) with
simulations. In addition to this quantitative criterion, we tried to
reproduce stop-and-go waves as a qualitative criterion. Stop-and-go waves are a
characteristic phenomenon for the single file movement. Only one
of three investigated models satisfies both criteria.
\end{abstract}
\section{Introduction}
The applications of pedestrians' dynamics range from the safety of
large events to the planning of towns with a view to pedestrian
comfort. Because of the computational effort involved with an
experimental analysis of the complex collective system of pedestrians'
behavior, computer simulations are run. Models continuous in space are
one possibility to describe this complex collective system.

In developing a model, we prefer to start with the simplest case: 
single lane movement. If the model is able to reproduce reality
quantitatively and qualitatively for that simple case, it is a good candidate for adaption 
to complex two-dimensional conditions.
 
Also in single file movement pedestrians interact in many ways
and not all factors, which have an effect on their behavior, are
known. Therefore, we follow three different modeling approaches in
this work. All of them underlie diverse concepts in the simulation of
human behavior.\\

This study is a continuation and enlargement of the validation introduced 
in \cite{Seyfried2006}. For validation, we introduce two criteria: On the one hand, the
relation between velocity and density has to be described
correctly. This requirement is fulfilled, if the modeled data
reproduce the fundamental diagram. On the other hand, we are aiming to
reproduce the appearance of collective effects. A characteristic
effect for the single file movement are stop-and-go waves as they are
observed in experiments \cite{Seyfried2009a}. We obtained all
empirical data from several experiments of the single file movement. 
There a corridor with circular guiding was built, so that it possessed 
periodic boundary conditions. The density was varied by increasing the 
number of the involved pedestrians. For more information about the
experimental set-up, see \cite{Seyfried2009a},\cite{Seyfried2005}.\\


\section{Spatially Continuous Models}
The first two models investigated are force based and the dynamics
are given by the following system of coupled differential equations
\begin{equation}
  m_i \frac{d\, v_i}{d \, t} = F_i \quad \mbox{with} \quad F_i= F_i^{drv} +
  F_i^{rep} \quad \mbox{and} \quad \frac{d\, x_i}{d\, t}
  = v_i 
\end{equation}
where $F_i$ is the force acting on pedestrian $i$. The mass is denoted
by $m_i$, the velocity by $v_i$ and the current position by
$x_i$. This approach derives from sociology \cite{Lewin1951}. Here
psychological forces define the movement of humans through their
living space. This approach is transferred to pedestrians dynamics and
so $F_i$ is split to a repulsive force $F_i^{rep}$ and a driven
force $F_i^{drv}$. In our case the driving force is defined as
\begin{equation}
  F_i^{drv} = \frac{v_i^0 - v_i}{\tau}, 
\end{equation}
where $v_i^0$ is the desired velocity of a pedestrian and $\tau$ their
reaction time.

The other model is event driven. A pedestrian can be in different
states. A change between these states is called event. The calculation
of the velocity of each pedestrian is straightforward and depends on
these states.


\subsection{Social Force Model}
The first spatially continuous model was developed by Helbing and
Moln\'{a}r \cite{Helbing1995} and has often been modified. According
to \cite{Helbing2000a} the repulsive force for pedestrian $i$ is
defined by
\begin{equation}
F_i^{rep}(t)  = \sum_{j\neq i}f_{ij}(x_i,x_j) + \xi_i(t) \quad \mbox{with} \quad
f_{ij}(x_i,x_j)  = -\partial_x A(\Delta\, x_{i,j}-D)^{-B},
\end{equation}
with $A=0.2, B=2, D=1\, [m], \tau=0.2\, [s]$ and $\Delta\, x_{i,j}$ is
the distance between pedestrians $i$ and $j$. The
fluctuation parameter $\xi_i(t)$ represents the noise of the
system. In two-dimensional scenes, this parameter is used to create
jammed states and lane formations \cite{Helbing2000a}. In this
study, we are predominantly interested in the modeled relation between
velocity and density for single file movement. Therefore the
fluctuation parameter has no influence and is ignored.

First tests of this model indicated that the forces are too strong,
leading to unrealistically high velocities. Due to this it is 
necessary to limit the velocity to $v_{max}$, as it is
done in \cite{Helbing1995}
\begin{equation}
v_i(t)=\left\lbrace
\begin{array}{cl}
  v_i(t), & \mbox{\quad if } |v_i(t)| \leq v_{max}\\
  v_{max}, & \mbox{\quad otherwise}
\end{array}\right. \enspace .
\end{equation}
In our simulation we set $v_{max}=1.34\, [\frac{m}{s}]$.

\subsection{Model with Foresight}
In this model pedestrians possess a degree of foresight, in addition
to the current state of a pedestrian at time step $t$. This approach
considers an extrapolation of the velocity to time step $t+s$. For it
\cite{Steffen2008} employs the linear relation between the velocity
and the distance of a pedestrian $i$ to the one in front $\Delta x_{i,i+1}(t)$.
\begin{equation}\label{v(d)}
  v_i(t)   =  a\,\Delta x_{i,i+1}(t) -b
\end{equation}
For $a=0.94\, [\frac{m}{s}]$ and $b=0.34\, [\frac{1}{s}]$ this
reproduces the empirical data. So  with (\ref{v(d)}) $v_i(t+s)$ can be
calculated from $\Delta x_{i,i+1}(t+s)$ which itself is a result of the
extrapolation of the current state
\begin{equation}
  \Delta x_{i,i+1}(t) + \Delta v(t)\, s = \Delta x_{i,i+1}(t+s)
\end{equation}
with $\Delta v(t) = v_{i+1}(t)\, s  -v_i(t)\, s$. Finally, the
repulsive force is defined as
\begin{equation}
  F_i^{rep}(t)=-\frac{v_i^0 -v_i(t+s)}{\tau} \enspace .
\end{equation}
Obviously the impact of the desired velocity $v_i^0$ in the
driven force is negated by the one in the repulsive
term. After some simulation time, the system reaches an equilibrium in
which all pedestrians walk with the same velocity. In order to spread
the values and keep the right relation between velocity and
density, we added a fluctuation parameter $\zeta_i(t)$. $\zeta_i(t)$
uniformly distributed in the interval $[-20, 20]$ reproduced the
scatter observed in the empirical data.

\subsection{Adaptive Velocity Model}
\begin{figure}[!h]
  \centering
  \subfigure[Demonstration of the parameters $d, st$ and $\beta$]
  {
    \includegraphics[height=4cm]{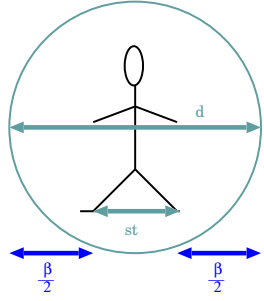}
    \label{sicherheitsabstand}
  }
  \hspace{1cm}
  \subfigure[Conception of the adaptive velocity]
  {
    \includegraphics[height=4cm]{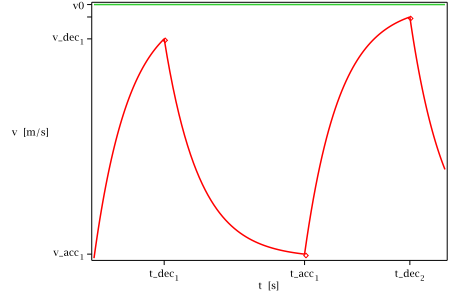}
    \label{adaptierteGeschwindigkeit}
  }
  \caption[]{Left: connection between the required space $d$, the step
    length $st$ and the safety distance $ \beta$. Right: The adaptive
    velocity with  acceleration until $t_{dec1}$, than deceleration
    until $t_{acc1}$, again acceleration until $t_{dec2}$ and so on.}
  \label{fig1}
\end{figure}

In this model pedestrians are treated as hard bodies with a diameter
$d_i$ \cite{Chraibi2009}. The diameter depends
linearly on the current velocity and is equal to
the step length $st$ in addition to the safety distance $ \beta$
\begin{equation}\label{d_i}
d_i(t)  =  e + f\, v_i(t) = st_i(t)+\beta_i(t) \enspace .
\end{equation}
Based on \cite{Weidmann1993} the step length is a linear
function of the current velocity with following parameters:
\begin{equation}\label{st_i}
st_i(t) = 0.235\, [m] + 0.302\, [s] \,v_i(t) \enspace .
\end{equation} 
$e$ and $f$ can be specified through empirical data and the inverse
relation of (\ref{v(d)}). Here $e$ is the required space for a stationary
pedestrian and $f$ affects the velocity term. For
$e=0.36\, [m]$ and $f=1.06\, [s]$ the last
equations (\ref{d_i}) and (\ref{st_i}) can be summarized to
\begin{equation}
\beta_i(t)=d_i(t)-st_i(t)=0.125 \, [m] +0.758\, [s] \, v_i(t)\enspace .
\end{equation}
By solving the differential equation 
\begin{equation}
\frac{dv}{dt}=F^{drv}=\frac{v^0 -v(t)}{\tau} \quad \Rightarrow \quad
v(t)=v^0+c\, \exp \left(-\frac{t}{\tau}\right), \mbox{for }c\in
\bbbr,
\end{equation} 
the velocity function is obtained. This is shown in
Fig. \ref{fig1} together with the parameters of the pedestrians'
movement.

A pedestrian is accelerating to their desired velocity $v_i^0$ until the
distance to the pedestrian in front is smaller than the safety
distance. From this time on, he/she is decelerating until the distance is
larger than the safety distance and so on.
Via $\Delta x_{i,i+1}, d_{i}$ and $\beta_i$ those events could be
defined: deceleration (\ref{Bremsen}) and acceleration
(\ref{Beschleunigen}). 

To ensure good performance for high densities, no events are explicitly
calculated. But in each time step, it is checked whether an event has
taken place and $t_{dec}, t_{acc}$ or $t_{coll}$ are
set to $t$ accordingly. The time step, $\Delta t$, of $0.05$ seconds is chosen, so
that a reaction time is automatically included. The
discrete time step could lead to configurations where overlapping
occurs. To guarantee volume exclusion, case (\ref{Stoppen}) is included, in which
the pedestrians are too close to each other and have to stop.  
\begin{eqnarray}
  t & = & t_{dec},\quad \mbox{if: } \quad \Delta x_{i,i+1}
  -0.5*\left(d_i(t)+d_{i+1}(t)\right) \leq 0\label{Bremsen}\\
  t & = & t_{acc},\quad \mbox{if: } \quad \Delta x_{i,i+1}
  -0.5*\left(d_i(t)+d_{i+1}(t)\right) > 0\label{Beschleunigen}\\
  t& = & t_{coll},\quad \mbox{if: } \quad \Delta x_{i,i+1}
  -0.5*\left(d_i(t)+d_{i+1}(t) \right)\leq -\beta_i(t)\label{Stoppen}
\end{eqnarray}


\section{Validation with Empirical Data}
\begin{figure}[!h]
    \centerline{
    \subfigure[Social force model]
    {
      \includegraphics[width=8cm]{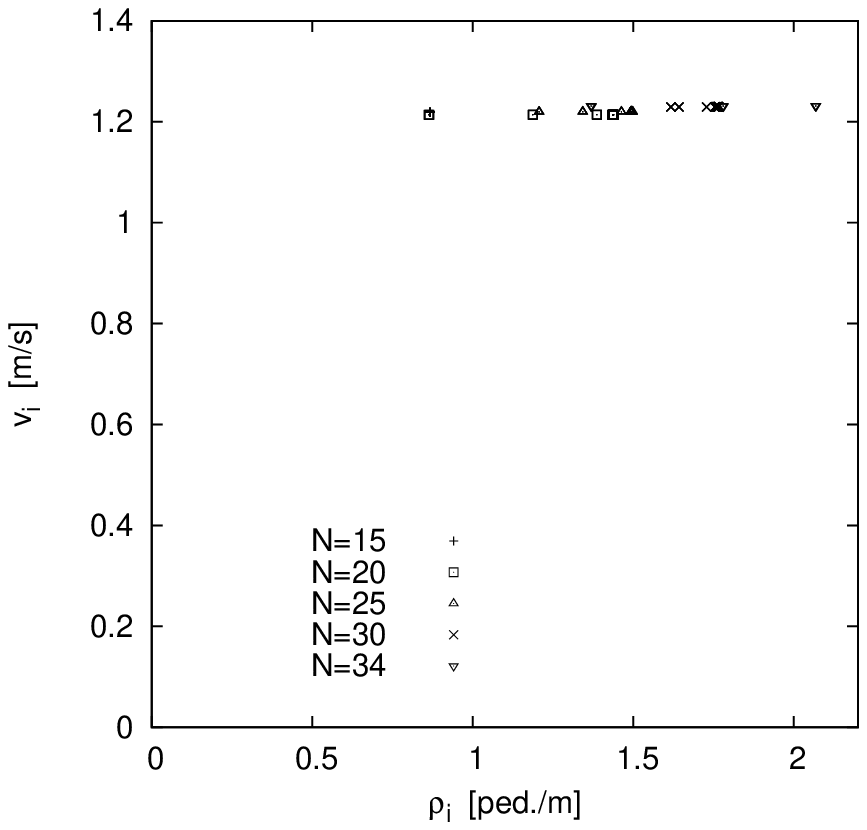}
      \label{}
    }
\hspace{-2cm}
    \subfigure[Model with foresight]
    {
      \includegraphics[width=8cm]{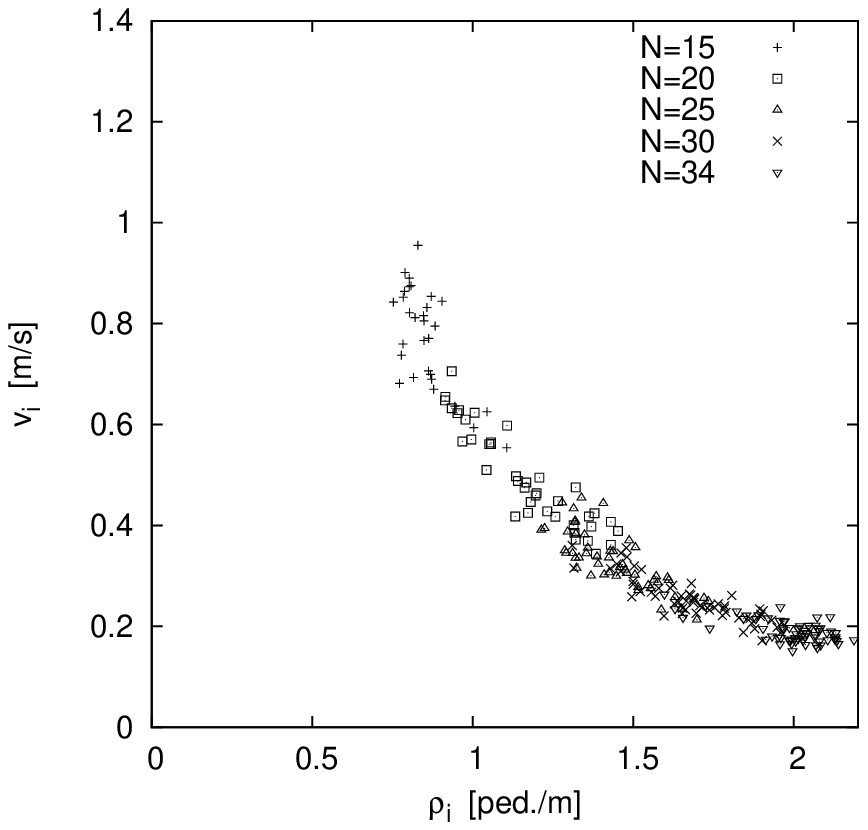}
      \label{}
    }
  }
  \centerline{
    \subfigure[Adaptive velocity model]
    {
      \includegraphics[width=8cm]{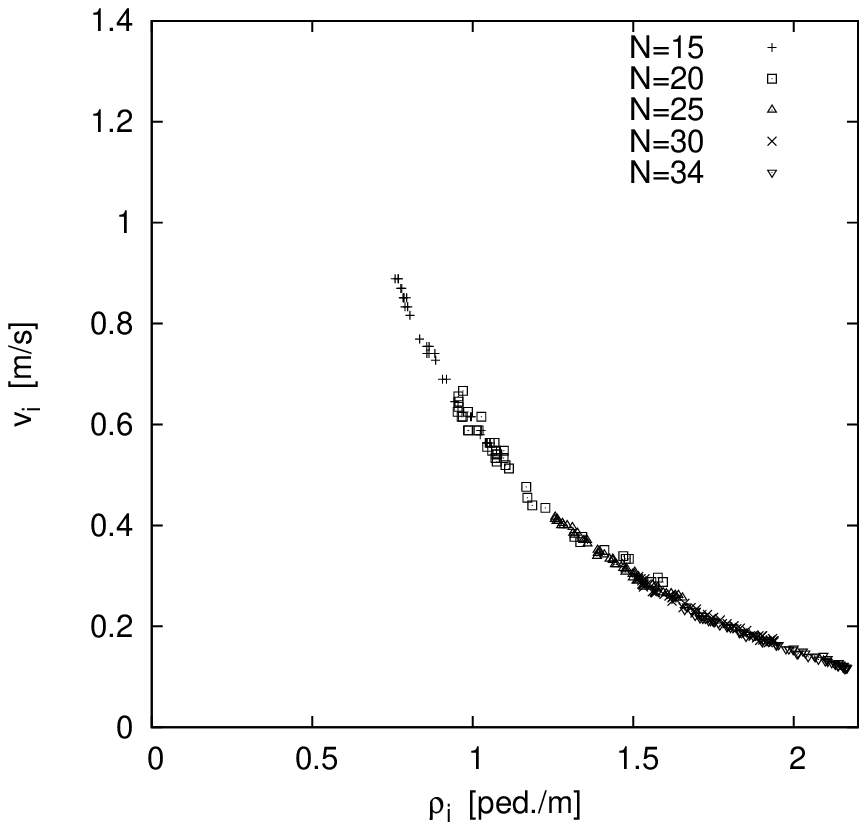}
      \label{}
    }
\hspace{-2cm}
    \subfigure[Empirical data]
    {
      \includegraphics[width=8cm]{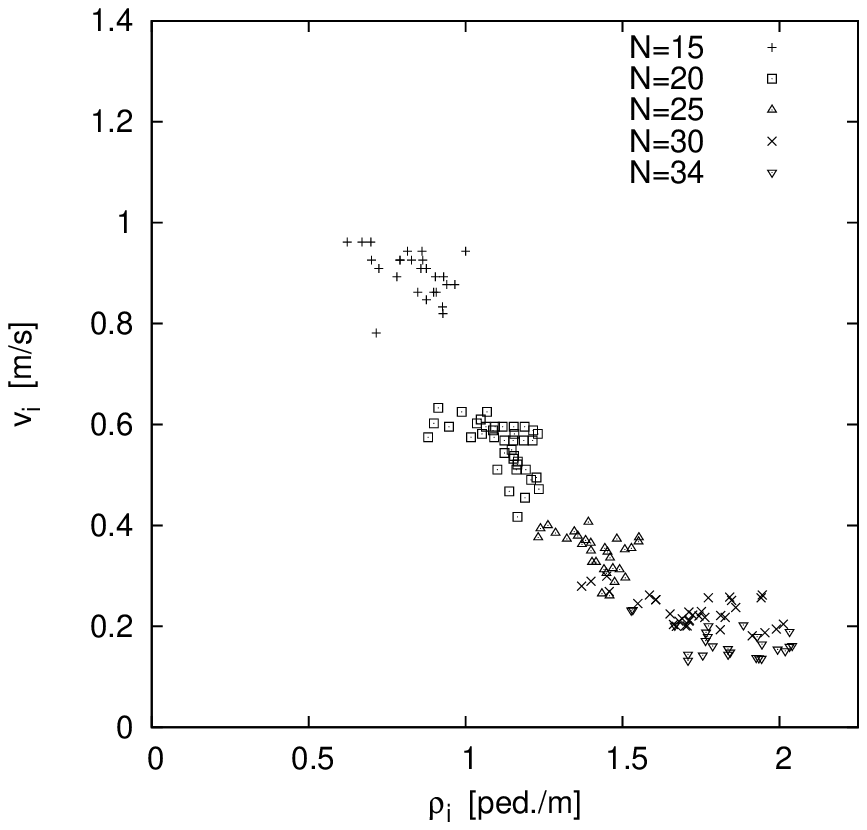}
      \label{}
    }
  }
  \caption[]{Validation of the modeled fundamental diagram with the
    empirical data (down right) for the single file movement.}
  \label{FD}
\end{figure}

For the comparison of the modeled and experimental data, it is
important to use the same method of measurement. \cite{Seyfried2009a}
shows that the results from different measurement methods vary
considerably. The velocity $v_i$ is calculated by the entrance and
exit times $t_i^{in}$ and $t_i^{out}$ to two meter section.
\begin{eqnarray}\label{v_man}
v_i & = & \frac{2\,  [m]}{(t_i^{out}-t_i^{in})[s]}.
\end{eqnarray}
To avoid discrete values of the density leading to a large scatter, we
define the density by
\begin{eqnarray}\label{dichte_man}
\rho(t) & = & \frac{\sum_{i=1}^N \Theta_i(t)}{2\, m},
\end{eqnarray}
where $\Theta_i(t)$ gives the fraction to which the space between pedestrian
$i$ and $i+1$ is inside the measured section, see
\cite{Seyfried2005}. $\rho_i$ is the mean value of all $\rho(t)$, where
$t$ is in the interval $[t_i^{in}, t_i^{out}]$.
We use the same method of measurement for the modeled and empirical
data. The fundamental diagrams are displayed in Fig. \ref{FD}, where
$N$ is the number of the pedestrians.

The velocities of the social force model are independent of the systems
density and nearly equal to the desired velocity $v_i=v_i^0 \sim
1.24\, [\frac{m}{s}]$. Additionally we observe a backward movement of the
pedestrians and pair formation. Because of these unrealistic phenomena are not observed in the
other models, we suggest that this is caused by the
combination of long-range forces and periodic boundary
conditions.

In contrast, the model with foresight results in a fundamental diagram
in good agreement with the empirical one. Through the fluctuation parameter, the
values of the velocities and densities vary as in the experimental
data.

We are satisfied with the results of the adaptive velocity model. For reducing
computing time, we also tested a linear adaptive velocity
function, leading to a $70 \%$ decrease in computing time
for $10000$ pedestrians. The fundamental diagram for
this linear adaptive velocity function is not shown, but also reproduces
the empirical one.\\

\section{Reproduction of Stop-and-Go Waves}
\begin{figure}[!h]
   \centerline{
    \subfigure[Model with foresight]
    {
      \includegraphics[height=3.75cm]{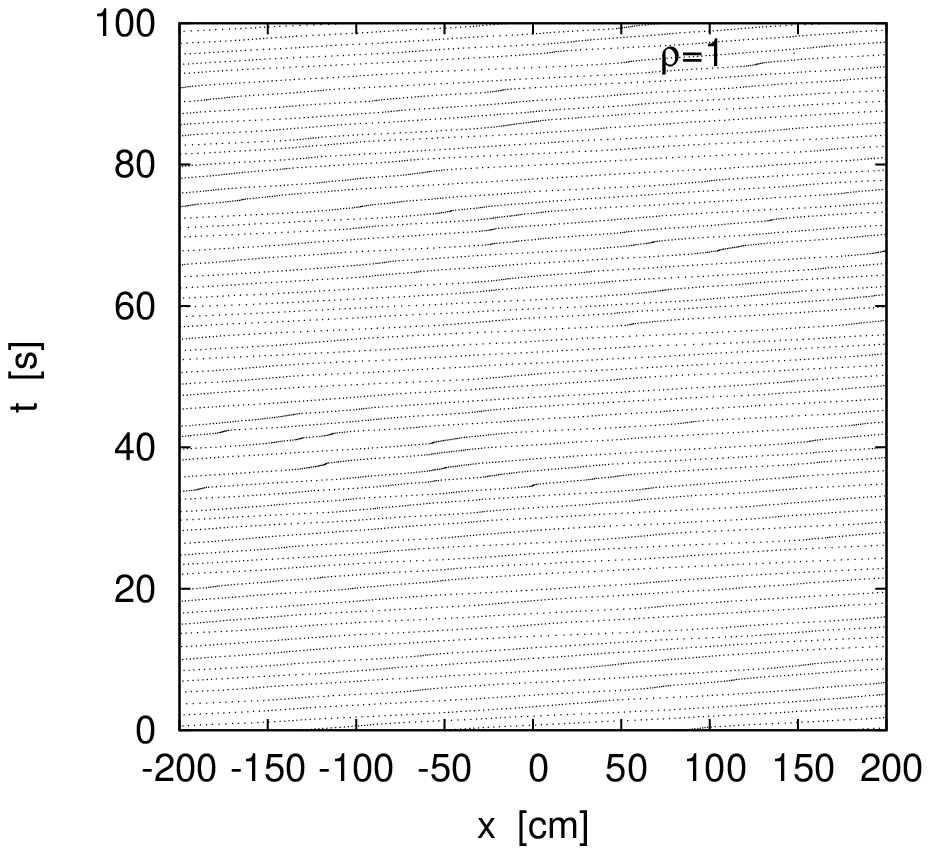}
      \label{Vor1}
    }
    \hspace{-1.8cm}
    \subfigure[Adaptive velocity model]
    {
      \includegraphics[height=3.75cm]{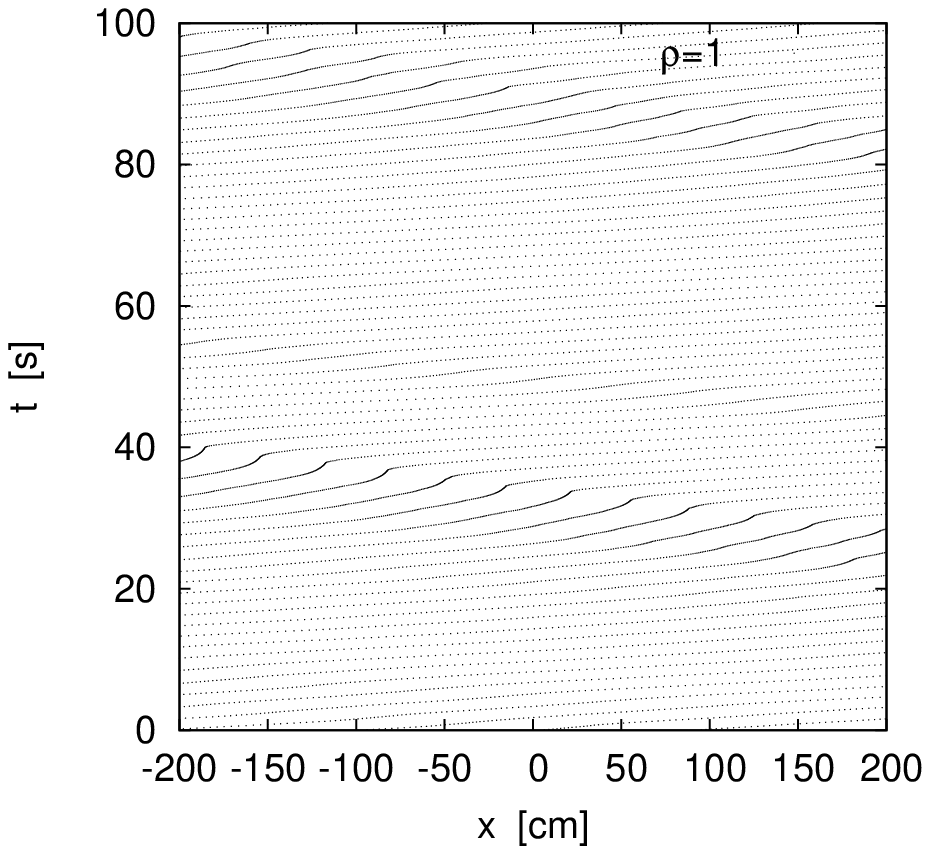}
      \label{ED1}
    }
    \hspace{-1.8cm}
    \subfigure[Experimental data]
    {
      \includegraphics[height=3.75cm]{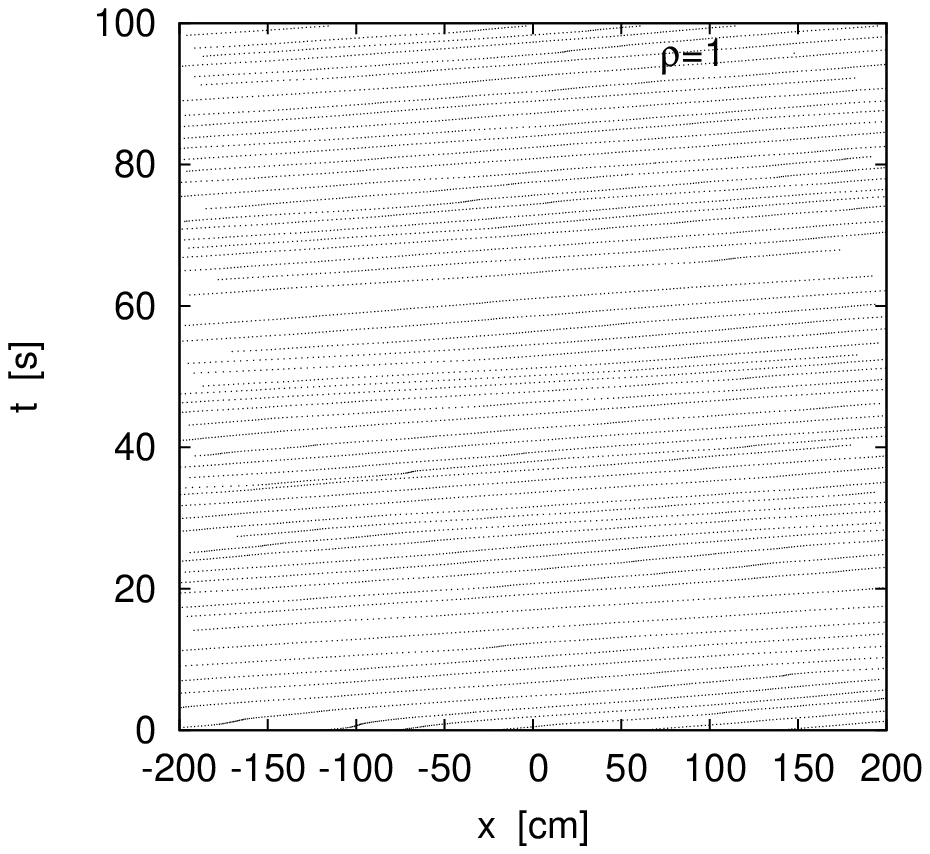}
      \label{emp1}
    }
  }
  \centerline{
    \subfigure[Model with foresight]
    {
      \includegraphics[height=3.75cm]{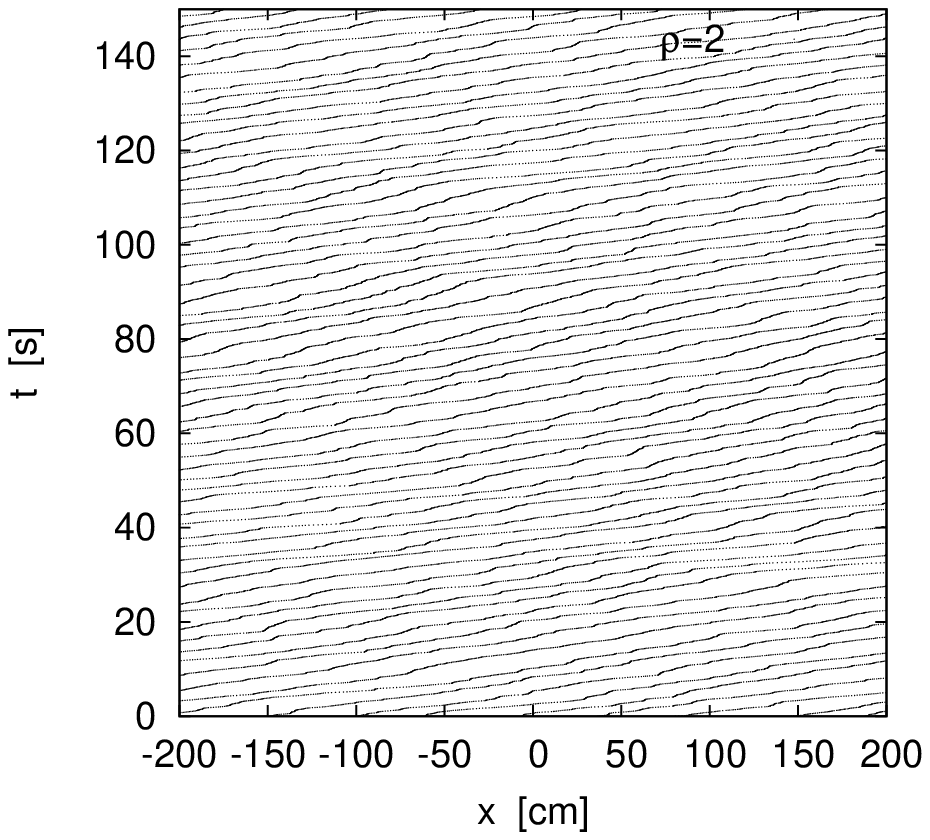}
      \label{Vor2}
    }
    \hspace{-1.8cm}
    \subfigure[Adaptive velocity model]
    {
      \includegraphics[height=3.75cm]{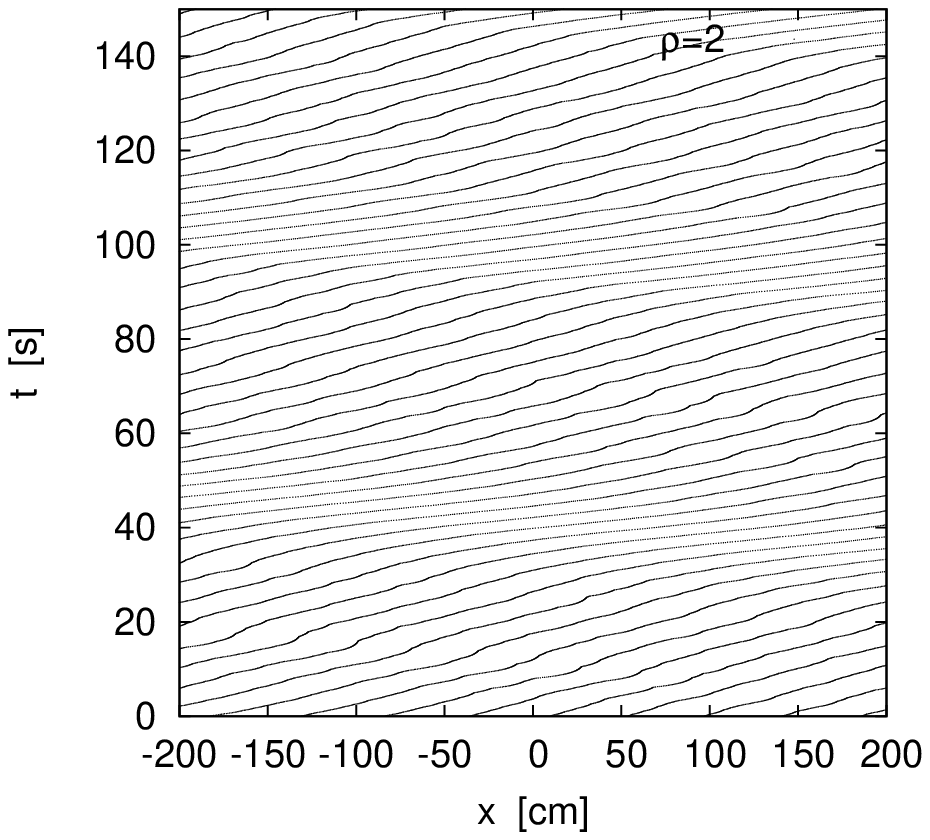}
      \label{ED2}
    }
    \hspace{-1.8cm}
    \subfigure[Experimental data]
    {
      \includegraphics[height=3.75cm]{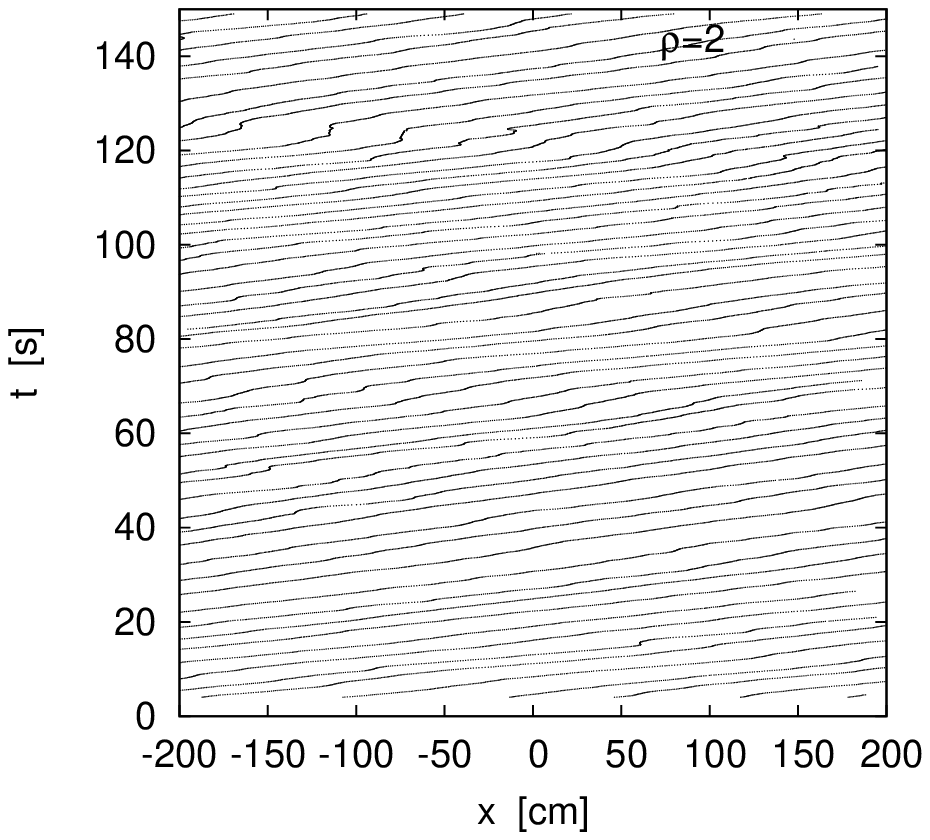}
      \label{emp2}
    }
  }
  \centerline{
    \subfigure[Model with foresight]
    {
      \includegraphics[height=3.75cm]{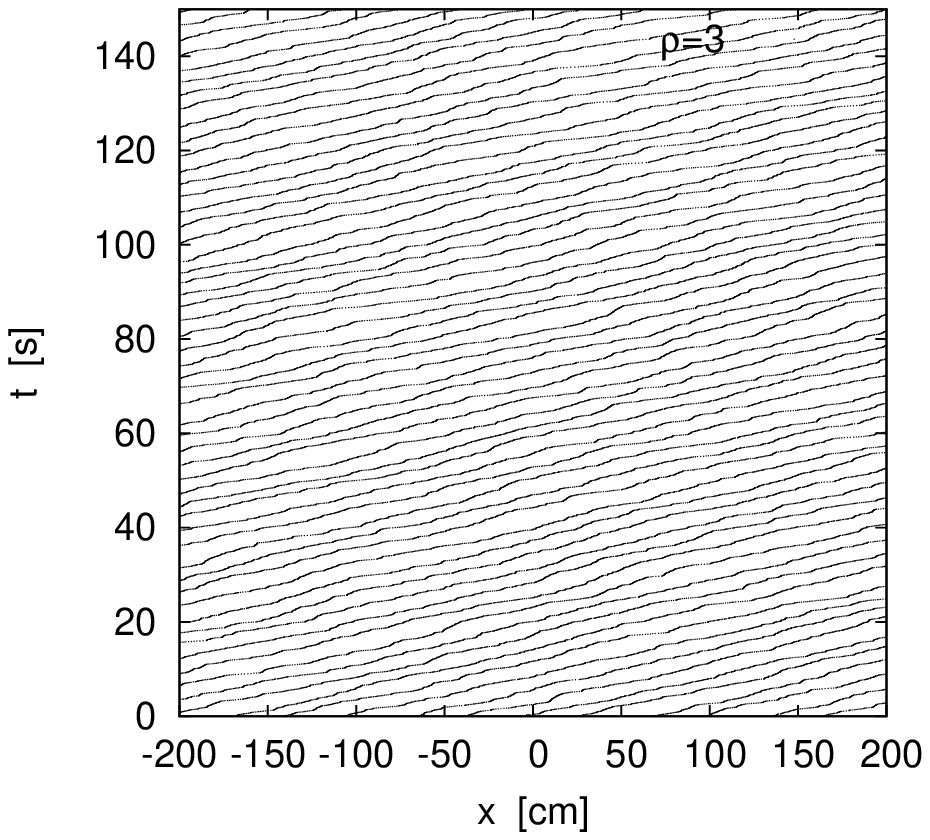}
      \label{Vor3}
    }
    \hspace{-1.8cm}
    \subfigure[Adaptive velocity model]
    {
      \includegraphics[height=3.75cm]{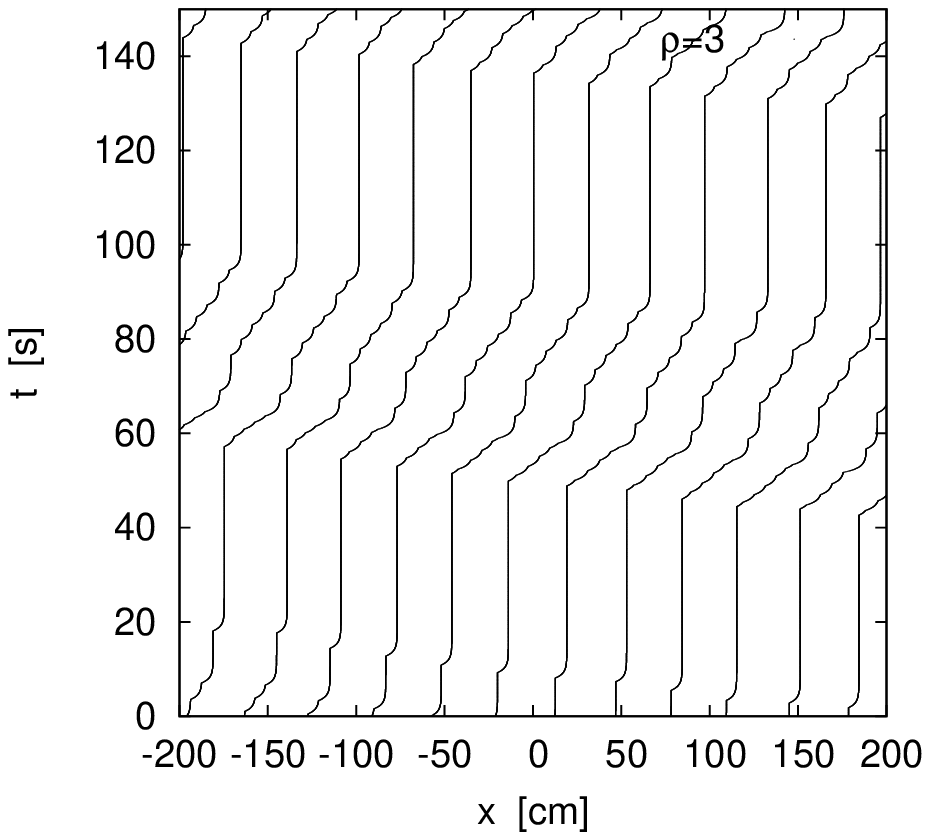}
      \label{ED3}
    }
    \hspace{-1.8cm}
    \subfigure[Experimental data]
    {
      \includegraphics[height=3.75cm]{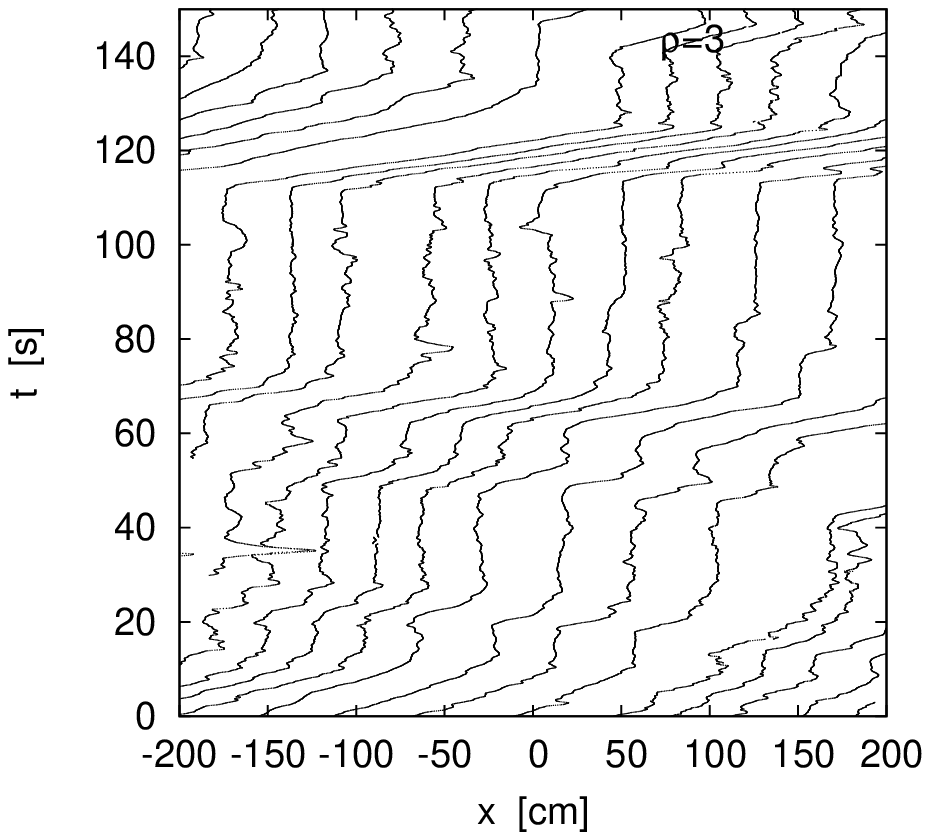}
      \label{emp3}
    }
  }
  \caption{Comparison of modeled and empirical trajectories for the single lane
    movement. The global density of the system is one, two or three persons
    per meter (from top to bottom).}
  \label{ges}
\end{figure}

During the experiments of the single file movement, we observed stop-and-go
waves at densities higher than two pedestrians per meter, see Fig. 5 in
\cite{Seyfried2009a}. Therefore, we compare the experimental
trajectories with the modeled ones for global densities of one, two
and three persons per meter. The results are shown in
Fig.\ref{ges}. Since the social force model is not able to satisfy the
criterion for the right relation between velocity and density, we omit this
model in this section.
Figure \ref{ges} shows the trajectories for global average densities
of one, two and three persons per meter. From left to right the data
of the model with foresight, the adaptive velocity model
and the experiment are shown.

In the experimental data, it is clearly visible that the trajectories
get unsteadier with increasing density. At a density of one person per
meter pedestrians stop for the first time. So a jam is generated. At a
density of two persons per meter stop-and-go waves pass through the
whole measurement range. At densities greater than three persons per
meter pedestrians can hardly move forward.

For the extraction of the empirical trajectories, the pedestrians' heads
were marked and tracked. Sometimes, there is a backward movement in
the empirical trajectories caused by self-dynamic of the pedestrians'
heads. This dynamic is not modeled and so the other trajectories have
no backward movement. This has to be accounted for in the comparison.

By adding the fluctuation parameter $\zeta_i(t) \in [-20,20]$ to the
model with foresight a good agreement with experiment is obtained for 
densities of one and two persons per meter. The irregularities
caused by this parameter are equal to the irregularities of the
pedestrians dynamic. Nevertheless, this does not suffice for stopping so that
stop-and-go waves appear, Fig.\ref{Vor2} and Fig.\ref{Vor3}.

With the adaptive velocity model stop-and-go waves already arise at a density
of one pedestrian per meter, something that is not seen in
experimental data. However, this model characterizes higher
densities well. So in comparison with Fig. \ref{ED3} and Fig. \ref{emp3} the stopping-phase 
of the modeled data seems to last for the same time as in the empirical data. But there 
are clearly differences in the acceleration phase, the adaptive velocity models
acceleration is much lower than seen in experiment.

Finally other studies of stop-and-go waves have to be carried out. The occurrence of 
this phenomena has to be clearly understood for further model modifications.
Therefore it is necessarry to measure e. g. the size of the stop-and-go wave at a fixed 
position. Unfortunately it is not possible to measure over a time interval, because 
the empirical trajectories are only available in a specific range of 4 meters.
\section{Conclusion}
The well-known and often used social force model is unable to
reproduce the fundamental diagram.
The model with foresight provides a good quantitative reproduction of
the fundamental diagram. However, it has to be modified further, so
that stop-and-go waves could be generated as well.
The model with adaptive velocities follows a simple and effective
event driven approach. With the included reaction time, it is possible
to create stop-and-go waves without unrealistic phenomena, like
overlapping or interpenetrating pedestrians.

All models are implemented in C and run on a simple PC. They were also tested for their
computing time in case of large system with upto 10000 pedestrians. The social force model 
offers a complexity level of $\mathcal{O}(N^2)$, whereas the other models only have a 
level of $\mathcal{O}(N)$. For this reason the social force model is not qualified for 
modeling such large systems. Both other models are able to do this, where the maximal
computing time is one sixth of the simulated time.

In the future, we plan to include steering of pedestrians. For these
models more criteria, like the reproduction of flow characteristics at
bottlenecks, are necessarry. Further we are trying to get a deeper insight into
to occurrence of stop-and-go waves.


%
\end{document}